\documentclass[aps,prb,twocolumn,hidepacs,hidekeys,superscriptaddress,10pt,floatfix]{revtex4}

\newcommand{\APP}{$\mathrm{AP \rightarrow P}$ }
\newcommand{\PAP}{$\mathrm{P \rightarrow AP}$ }

\usepackage[normalem]{ulem}
\usepackage{amsmath}
\usepackage{amssymb}
\usepackage{latexsym}
\usepackage{indentfirst}
\usepackage{graphicx}
\usepackage{color}
\usepackage{bm}
\usepackage{multirow}

\begin{document}
\title{Spin-transfer assisted thermally activated switching distributions in perpendicularly magnetized spin valve nanopillars}

\author{D.~B. Gopman}
\email{daniel.gopman@physics.nyu.edu}
\affiliation{Department of Physics, New York University, New
             York, NY 10003, USA}
\author{D. Bedau}
\affiliation{HGST San Jose Research Center,
             San Jose, CA 95135 USA}
\author{S. Mangin}
\affiliation{Institut Jean Lamour, UMR CNRS 7198 –Universit\'{e} de Lorraine, Nancy, France}
\author{E.~E. Fullerton}
\affiliation{CMRR, University of California at San Diego,
             La Jolla, CA 92093, USA}
\author{J.~A. Katine}
\affiliation{HGST San Jose Research Center,
             San Jose, CA 95135 USA}
\author{A.~D. Kent}
\affiliation{Department of Physics, New York University, New
             York, NY 10003, USA}
             
\begin{abstract}
We present switching field distributions of spin-transfer assisted magnetization reversal in perpendicularly magnetized Co/Ni multilayer
spin-valve nanopillars at room temperature. Switching field measurements of the Co/Ni free layer of spin-valve nanopillars with a 50~nm$\times$~300~nm ellipse cross section were conducted as a function of current. The validity of a model that assumes a spin-current dependent effective barrier for thermally activated reversal is tested by measuring switching field distributions under applied direct currents. We show that the switching field distributions deviate significantly from the double exponential shape predicted by the effective barrier model, beginning 
at applied currents as low as half of the zero field critical current. Barrier heights extracted from switching field distributions for currents below this threshold are a monotonic function of the current. However, the thermally-induced switching model breaks down for currents exceeding the critical threshold.
\end{abstract}


\maketitle
\section{Introduction \label{Intro}}
Spin transfer driven magnetization reversal is of great fundamental interest and has a direct impact on magnetic information storage technologies.\cite{Katine2008} Nanostructures with perpendicular magnetic anisotropy are of particular importance to storage applications.\cite{Kent2010,Ikeda2010,Brataas2012} The all-perpendicular geometry yields reduced critical currents $ \text{I}_\text{c}$, high uniaxial symmetry and high spin-torque switching efficiency, i.e., a small ratio of the critical current to energy barrier $ \text{I}_\text{c} $/U.\cite{Sun2000,Mangin2006,Mangin2009} Implementation of competitive spin transfer devices requires low critical currents while maintaining sufficient thermal stability to suppress thermally-activated switching between magnetization configurations. 

Understanding the thermal stability of a nanomagnet under spin-transfer torque (STT) is of critical importance to predicting device performance, particularly in the sub-threshold current drive regime in which devices can still switch by thermal activation. In the absence of STT, the probability that at finite temperature a nanomagnet's direction of magnetization switches in an applied magnetic field is expected to follow a simple model of thermal activation over an energy barrier.\cite{Neel1948, Brown1963} A widely used recent model predicts that spin-transfer torques lead to a spin-current dependent effective energy barrier for thermally assisted transitions.\cite{Li2004, Sun2006} This model predicts that a nanomagnet under STT reaches a new steady state that corresponds to an equilibrium distribution over magnetic configurations with an effective potential energy landscape that is modified by the current. The predictions from this model were investigated numerically using Fokker-Planck calculations\cite{Apalkov2005} and empirically using dwell-time measurements of in-plane magnetized nanopillar devices.\cite{Krivorotov2004} 

Recent spin-torque switching studies in perpendicularly magnetized nanopillar spin-valves have applied this model. Experimentally obtained energy barrier heights were shown to be much lower than the uniaxial barrier height determined by the entire magnetic free layer volume.\cite{Bedau2010b} Nevertheless, the switching appears well described by thermally overcoming a single energy barrier, whose height is related to an excited magnetic subvolume in the free layer element.\cite{Sun2011} Standard measurements probing the effects of spin-torques on switching - current-field state diagrams and measurements of the Stoner-Wohlfarth astroid - also appear to agree with the simple effective barrier model.\cite{Mangin2006, LeGall2012, Worledge2011, Henry2009} 

In order to further test the validity of this effective model under STT, it is important to probe the thermal switching behavior of a spin-torque driven nanomagnet. We focus on the magnetization reversal characteristics of the Co-Ni free layer (FL) element in all metallic spin-valve (SV) nanopillars with a perpendicularly magnetized polarizing reference layer (RL) composed of Co-Ni and Co-Pt. Spin valves with both the polarizer and the free layer having perpendicular magnetic anisotropy (PMA) are a uniaxial model system, in which all of the contributions (internal and external fields, anisotropy axis, and spin-current axis) are nearly aligned perpendicular to the film plane. Co-Ni multilayered films show high PMA, significant spin-polarization and low Gilbert damping compared to other PMA systems (Co/Pt, Co/Pd, FePt).\cite{Daalderop1992, Denbroeder1992, Beaujour2007, Girod2009, Beaujour2011} Furthermore, the all-metal system allows us to generate current densities higher than those possible in magnetic tunnel junctions.\cite{Bedau2010a} 

In this paper we focus on the influence of STT on thermally-assisted reversal. After a brief description of the spin valves studied here, we demonstrate a variety of methods to probe the thermally activated reversal characteristic of a nanomagnet. We begin by introducing standard measurements probing the effects of spin-torques on switching - the current-field state diagram. We then present measurements of the coercivity versus field-sweep rate under several direct currents to probe changes in the spin-current-dependent effective energy barrier height. Finally, we focus on statistical measurements of the switching field under finite direct currents. Using a switching field model for thermal activation over a single energy barrier, we extract the effective barrier height from switching field distribution measurements at each applied current in order to monitor the evolution of the effective barrier height with current. We have obtained over 5,000 switching events at each applied current, which allows us to sample a relatively large number of the statistically rare events at the distribution tails. These statistically rare events are indicators of where deviations from an equilibrium (effective barrier) model first emerge. We use a Gauss quantile plot of the switching field distributions to highlight the data at the distribution tails and demonstrate deviations of our data from the model for the top one percent of the switching probability for currents below the zero-field critical switching current $I_c$. Furthermore, significant deviations from the equilibrium model emerge after exceeding this current threshold. 
\section{Device Fabrication and Electrical Measurements \label{Materials}}
The Co/Ni nanopillars studied here are part of an all-perpendicular spin valve device. Details on materials and sample preparation have been reported previously.\cite{Mangin2006} The magnetic multilayered structure consists of a Pt(3)\slash [Co(0.25)\slash Pt(0.52)]$\times$4\slash Co(0.25)\slash [Ni(0.6)\slash Co(0.1)]$\times$2 hard reference layer and a [Co(0.1)\slash Ni(0.6)]$\times$2\slash Co(0.2)\slash Pt(3) free layer separated by a 4~nm Cu spacer layer and patterned into $\mathrm{50 \times 300 \, nm^2}$ ellipse-shaped nanopillars by a process that combines e-beam and optical lithography. Measurements were taken at room temperature and with fields applied within 3 degrees of the free layer easy axis. The reference layer magnetization switches for an applied field close to 1~T. Since no fields greater than 0.3~T are applied during the measurements, the reference layer is expected to remain fixed and pointing along the direction of negative magnetic fields, unless otherwise specified.

The magnetization of the free layer is probed indirectly with four-probe measurements of the spin-valve magnetoresistance. For the experiments under constant dc current, we used differential resistance measurements under a 10~kHz excitation current of $I_{ac}=100\; \mu $A~rms (the room temperature, zero-field critical switching current, $ I_c \approx~5~\text{mA} \gg I_{ac} $) using standard lock-in techniques. 
\section{State Diagram: Effect of spin-transfer torques on coercivity \label{StateDiag}}
\begin{figure}[b]
  \begin{center}
    \includegraphics[width=3.375 in,
    keepaspectratio=True]
    {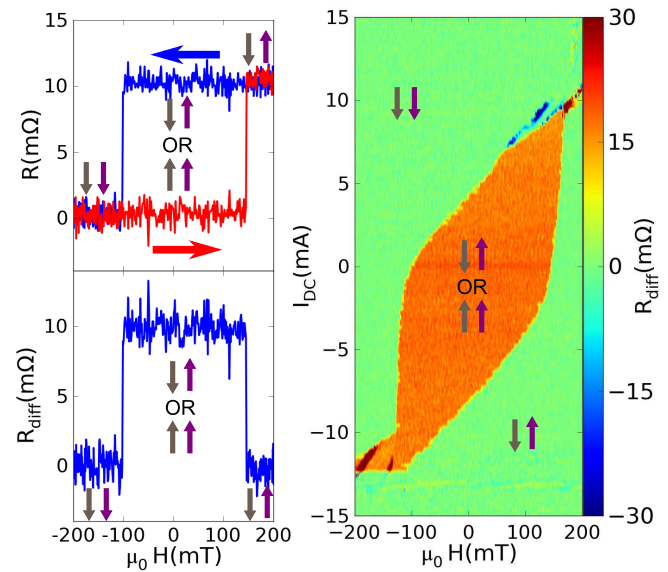}
  \end{center}
  \caption{\label{fig:StateDiagram}  Experimental state diagram of a 50~$\times$~300~$\mathrm{nm^2}$ ellipse spin-valve device: (a) red and blue curves show the increasing and decreasing branches of the resistance versus perpendicular applied field hysteresis loop. (R is the resistance deviations from $\mathrm{R_0=3.218 \, \Omega}$.)  (b) We subtract the resistances of the decreasing branch from the increasing branch $\mathrm{\Delta R (H) = R_{dec}(H) - R_{inc}(H)}$. (c) A series of resistance difference traces from hysteresis loops at applied currents $\mathrm{|I_{DC}| \leq 15 \, mA}$ is used to generate an interpolated density map, which defines the state diagram. This density map corresponds to the states available to the spin-valve device: green regions indicate only one state (anti-parallel or parallel), which orange regions represent an area of bistability. Vertical arrows illustrate the magnetization orientations of the two layers in each region.}
\end{figure}

The current-field state diagram investigates the stability of different spin-valve states under STT from electric currents as well as applied magnetic fields. This diagram reveals the regions of applied fields and currents that exhibit only an anti-parallel (AP) or parallel (P) state as well as bistable regions where either AP or P states can be stabilized. Figure~\ref{fig:StateDiagram} illustrates the state diagram of one of our $\mathrm{50 \times 300 \, nm^2}$ SV devices alongside the method for generating the diagram from a series of field hysteresis loops under many different applied currents. Figure~\ref{fig:StateDiagram}(a) presents a resistance versus perpendicular applied field hysteresis loop with a decreasing field (\APP) and an increasing field (\PAP) branch. The resistance difference between the \APP and \PAP branches at each applied field value is plotted in Fig.~\ref{fig:StateDiagram}(b), revealing field ranges with a low resistance difference, indicating that only one state (AP or P) is stable. There is also a region with a substantial resistance difference, indicating a region of bistability. The set of resistance difference traces obtained from field hysteresis loops at a series of currents is interpolated into a density map that is presented as the state diagram in Fig.~\ref{fig:StateDiagram}(c).
A trend line can be seen along the perimeter between the orange (bistable) region and the green (AP or P) regions. The linear dependence of the switching currents with applied field in a region surrounding zero field can be understood within a modified N\'eel-Brown law in which the spin-current modifies the effective barrier separating AP and P states. This result is consistent with finite temperature calculations of the current-field evolution of the state diagram.\cite{Zhu2008} We define the zero-field critical switching current as the room-temperature switching current at zero field, $\mathrm{|I_c|(H=0)}$. For \APP the critical current is 5~mA and for \PAP it is -7~mA. The sudden increase in slope $\mathrm{dI_c/dH}$ for fields $\mathrm{|\mu_0H|} \gtrsim 100$~mT cannot be understood by a modified N\'eel-Brown law, but tilts of the applied field relative to the uniaxial axis and higher-order terms in the uniaxial potential energy landscape (e.g. $\sin ^{2n} \theta, n \geq 2$) may be important for the origin of the deviations from the predicted linear dependence.\cite{LeGall2012}
\section{Field-sweep rate measurements \label{RateMeasurements}}
\begin{figure}[htb]
  \begin{center}
    \includegraphics[width=3.375in,
    keepaspectratio=True]
    {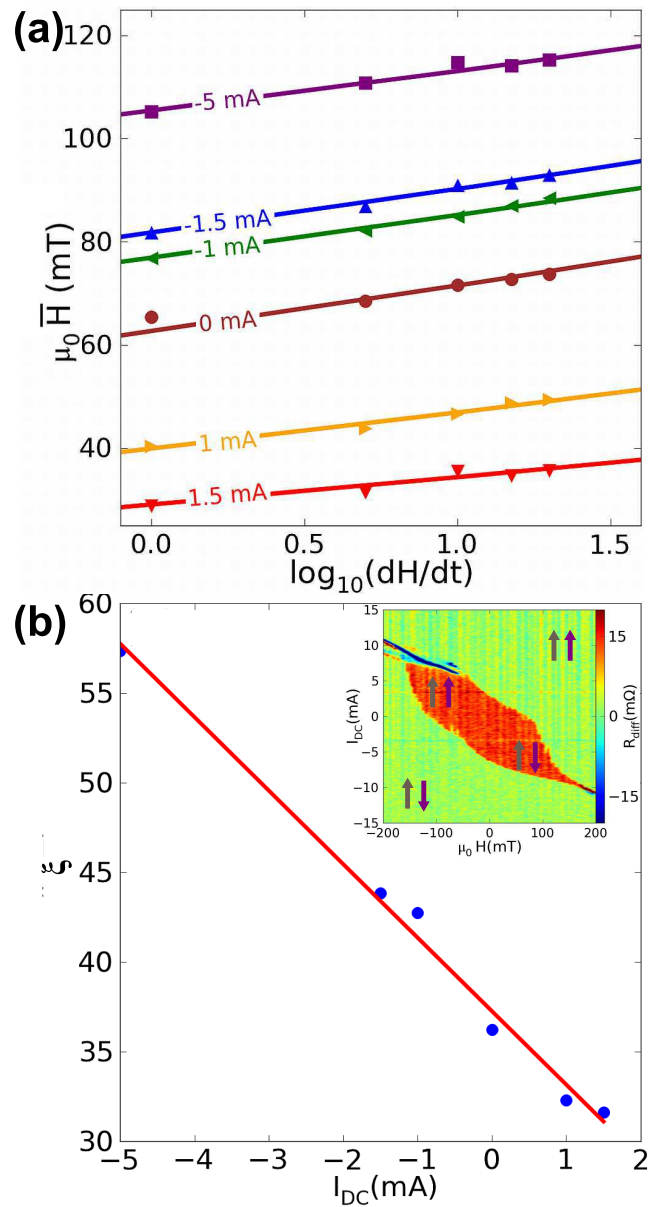}
  \end{center}
  \caption{\label{fig:FieldRateDependence} (a) Dependence of the mean switching field ($\mathrm{\mu_0\overline{H}}$) of the \APP transition on field sweep rate (dH/dt) for a second 50~$\times$~300~$\mathrm{nm^2}$ ellipse spin-valve device of the same composition with somewhat different characteristics subjected to several applied dc currents.  (b) Evolution of the extracted effective thermal stability parameter ($\xi$) versus dc current ($\mathrm{I_{DC}}$). Thermal stability factor obtained from best-fit lines (Eq.~\ref{meanswitch}) to data in (a). Inset to (b) showing the state diagram for this second device.}
\end{figure}

We can probe changes in the thermal stability of a nanomagnet under STT through variable field sweep rate measurements. The evolution of the mean switching field versus field sweep rate is sensitive to the nanomagnet's thermal stability factor $\xi = E_0 / k_BT$, where $E_0$ is the barrier height at zero applied field, $k_B$ is the Boltzmann constant and $T = 300$~K for our measurements. We assume an Arrhenius-type law for thermal activation, $\Gamma(H) = \Gamma_0 \exp(-\xi \varepsilon ^ \eta)$, where $\varepsilon = (1 - H/H_{c0})$ and $\eta = 1.5$\cite{eta,Victora1989,Coffey1995} determine the scaling of the thermal stability with field, $H_{c0}$ is the switching field at zero temperature and we assume $\Gamma_0 = 1$~GHz. Then, the cumulative probability that the nanomagnet does not switch under a magnetic field ramped linearly in time ($dH/dt = v =$~const.) from zero up to a field H has the form of a double exponential:\begin{eqnarray}
&& P_{NS}(H) = \exp \left( -\int_0^H \Gamma(H') / v dH'\right).
\label{doubleexponent}
\end{eqnarray}
This thermal activation expression models the experimentally obtained switching field distributions that we will introduce further below, but also yields an approximate expression for the mean switching field for a given field ramping rate $v$:\begin{eqnarray}
\overline{H}(v) \cong H_{c0} \left( 1 - \left[ \xi \ln \left( \frac{ \Gamma _0 H_{c0} }{ \eta v \xi \varepsilon ^{ \eta - 1} } \right)  \right]^{1 / \eta} \right).
\label{meanswitch}
\end{eqnarray}
We use the above expression to fit the evolution of the mean switching field with sweep-rate for a series of direct currents in order to determine the evolution of the thermal stability parameter $\xi$ with current.

We have conducted statistical switching field measurements under STT of multiple spin-valve devices, presenting similar behavior although with varying thermal stability. To better highlight the features general to these devices, we will present the results in this section on variable field-sweep rate measurements on a second spin-valve device of the same size and composition with somewhat different characteristics than the device studied in Fig.~\ref{fig:StateDiagram}. Figure~\ref{fig:FieldRateDependence}(a) shows the evolution of the mean switching field for the \APP transition versus the logarithm of the field sweep-rate along with best-fit trendlines fitting the data from Eq.~\ref{meanswitch}. From these trend lines, we have extracted the thermal stability $\xi$ for each applied current. The evolution of the thermal stability versus applied current is plotted in Fig.~\ref{fig:FieldRateDependence}(b), with the state diagram for this device provided in the inset as comparison to the device studied in Fig~\ref{fig:StateDiagram}. Applying a linear fit to the dataset consistent with an effective barrier model,\cite{Li2004} we extrapolate a zero-temperature critical switching current $\mathrm{I_{c0} = 9\, mA}$. This value is consistent with the closing of the bistable region of the state diagram plotted in the inset. 

In Fig.~\ref{fig:FieldRateDependence}, we presented results on the mean switching field of an \APP transition that agrees with the predictions of an effective barrier model under STT. We will proceed further to offer a more rigorous test to the model through statistical measurements distributions under STT. We will investigate the entire switching field distribution to test the extent to which the data agrees with the switching field model at the distribution tails. In the following section, we demonstrate that plotting the switching field distributions on a Gauss quantile scale permits us to better assess the quality of fit to our statistical data at the rare events comprising the tails of the distributions.

\section{Quantile scale plotting of the switching field distributions \label{sec:QuantileGauss}}
\begin{figure}[htb]
  \begin{center}
    \includegraphics[width=3.375 in,
    keepaspectratio=True]
    {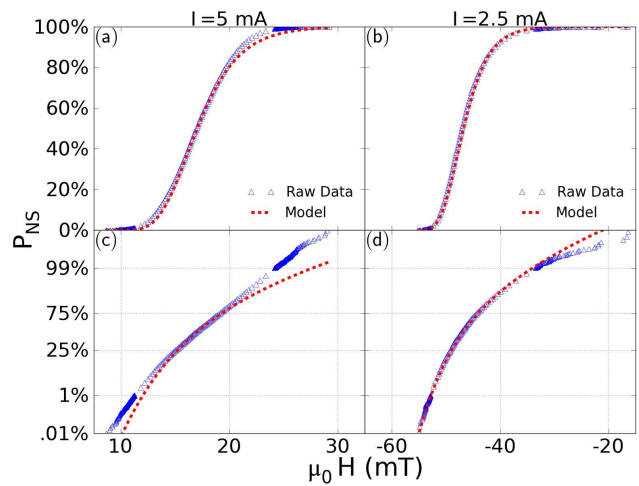}
  \end{center}
  \caption{\label{fig:Composite}  Switching field distributions for $\mathrm{AP \rightarrow P}$ transition under constant dc currents (2.5~mA~(b,d) \& 5~mA~(a,c)). Subfigures (a,b) are plotted on a linear y-scale and (c,d) are plotted on a Gauss quantile y-scale rescaled by $Y = \sqrt{2} \, \mathrm{inverf}(2y - 1)$ to magnify the datasets at the tails of the distributions. This plotting scheme highlights deviations of the data from the thermal activation model at the distribution tails.}
\end{figure}
Figure~\ref{fig:Composite} shows the distributions of switching fields for over 5000 switching field events, $\mathrm{AP \rightarrow P}$, under direct currents of 2.5~mA~(b,d) and 5~mA~(a,c), for the SV device first introduced in Fig.~\ref{fig:StateDiagram}. The switching field distributions in the top half (Figs.~\ref{fig:Composite}(a,b)) are plotted on linear axes for both applied field $\mu_0H$ and for the cumulative non-switching probability $\mathrm{P_{NS}}$, while in the bottom half (Figs.~\ref{fig:Composite}(c,d)) we plot $\mathrm{P_{NS}}$ on a Gaussian quantile scale. This rescaling permits us to qualitatively assess (a) the double-exponential character of thermal activation (Eq.~\ref{doubleexponent} giving rise to the asymmetric shape of the distribution  (e.g. non-Gauss) and (b) the quality of fitting the rare events at the tail of our distribution by stretching the y-axis around where the tails of the distribution is normally condensed. We map the y-axis representing $\mathrm{P_{NS}}$ onto a Gauss-quantile scale using the following rescaling of the y-axis:\begin{eqnarray}
&& Y = \sqrt{2} \, \mathrm{inverf}(2y - 1),
\label{QuantileGauss}
\end{eqnarray}
in which inverf is the inverse error function. For a normal (Gaussian) distribution, the data will collapse onto a line whose slope is equal to the inverse of $\sigma$, the standard deviation of the mean. The symmetric shape of a Gauss distribution is inconsistent with switching field distributions, in which thermal activation skews the distribution toward lower fields. This is why our switching field data curves away from an imaginary tangent line at the median ($P_{NS} = 0.5$) due to the double-exponential character of thermal activation. 

Figures~\ref{fig:Composite}(a,c) clearly show that the switching field distribution under a 5~mA current (open blue triangles) deviates sharply from the thermal activation model (dashed red line). The deviations appear over a sufficiently large region of the distribution that it is visible even on the linearly scaled axis in (a). However, when we compare the distributions under a 2.5~mA current in Figs.~\ref{fig:Composite}(b,d), the deviations are too subtle to ascertain from the linearly scaled plot in the top right corner. Once we plot our data on the Gauss quantile-scaled y-axis, disagreement between data and model at the tails of this distribution becomes clear. This result shows that deep statistical measurements of the switching field reveals problems at high current density and the gradual onset of deviations from our model at the distribution tails.

\section{Testing the model: Switching field distributions under constant dc currents \label{TestModel}}
In this section, we will test the scope of the ``modified barrier'' model by conducting deep statistical measurements of the switching field over a wide range of dc currents. Assuming thermal activation over a single energy barrier, at fixed temperatures the cumulative probability to remain in a metastable magnetization state under finite field, $\mu_0H$, is given by the double-exponential expression in Eq.~\ref{doubleexponent}. Our previous switching field studies taken in zero dc current were consistent with the single barrier model and serve as a baseline from which to compare switching distributions with finite dc currents.\cite{Gopman2012APL,Gopman2013} In order to test the effective barrier model, we will permit the thermal stability and zero-temperature coercive field to vary in Eq.~\ref{doubleexponent}, as spin-currents may modify the thermal stability $\xi$ as well as the effective anisotropy field $H_{c0}$ for switching.

\begin{figure}[t]
  \begin{center}
    \includegraphics[width=3.375in,
    keepaspectratio=True]
    {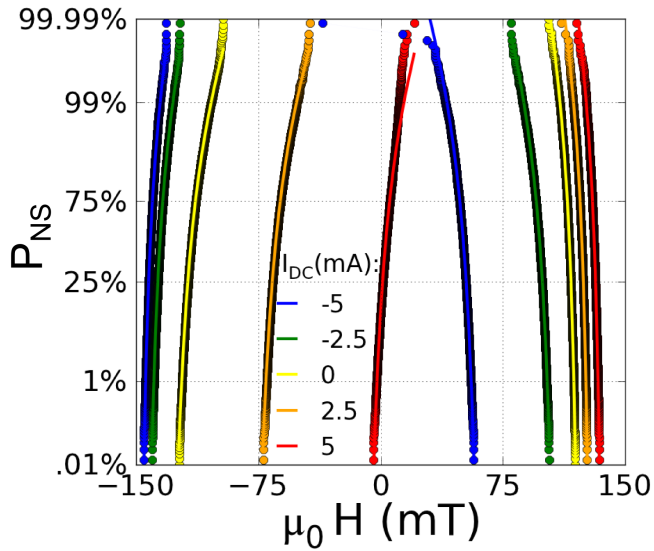}
  \end{center}
  \caption{\label{fig:SFDs} Switching field distributions for 5,000 events under direct currents ($\mathrm{I_{DC} = 0, \pm 2.5 , \pm 5 \, mA}$) plotted on a Gauss quantile scale. $\mathrm{AP \rightarrow P}$ ($\mathrm{P \rightarrow AP}$) distributions fall to the left (right) of the imaginary line at $\mu_0H = 0$. }
\end{figure}

Figure~\ref{fig:SFDs} shows the distributions of switching fields for 5000 switching events, both $\mathrm{AP \rightarrow P}$ and $\mathrm{P \rightarrow AP}$ , under direct currents of +/-5~mA, +/-2.5~mA and 0~mA for the first device shown in Fig.~\ref{fig:StateDiagram}. Curves to the right of zero magnetic field correspond to $\mathrm{P \rightarrow AP}$ transitions and to the left correspond to the $\mathrm{AP \rightarrow P}$ transitions. The switching field distributions for the 5000 events at each current are plotted on a Gaussian quantile scale, introduced previously in Section~\ref{sec:QuantileGauss} to assess the quality of fitting the rare events at the tails of our distributions.

For the single barrier model we apply to each dataset, we observe qualitatively good agreement of the fits to the majority of our measured switching field data. Furthermore, the centers of the distributions are shifted according to the applied currents. This behavior is consistent with the evolution of the switching field with applied current presented earlier in the State Diagram boundaries. However, we note that the first 1~\% of $\mathrm{AP \rightarrow P}$ transitions for $I_{DC} = 2.5, 5$~mA occur with lower probability than the model would predict, extrapolating outward from the median. We note that deviations from the model at 5~mA also coincide with the onset of switching for the $\mathrm{AP \rightarrow P}$ transition at zero applied field (see Fig.~\ref{fig:StateDiagram}).

\begin{figure}[b]
  \begin{center}
    \includegraphics[width=3.375in,
    keepaspectratio=True]
    {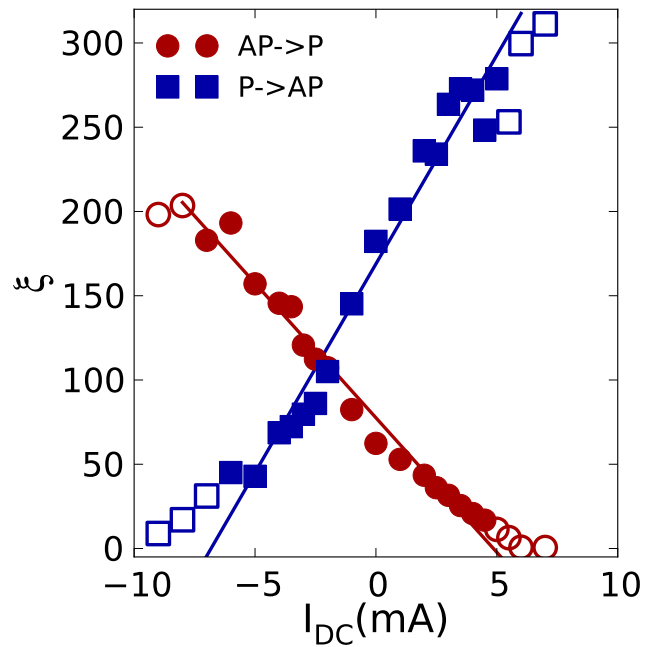}
  \end{center}
  \caption{\label{fig:Modified_Barrier} Effective thermal stability $\xi$ versus direct current $\mathrm{I_{DC}}$ for $\mathrm{|I_{DC}| < 9 \, mA}$. Thermal stability $\xi = E_0 / k_B T$) for $\mathrm{AP \rightarrow P}$ ($\mathrm{ P \rightarrow AP }$) transitions are referenced to 300~K and plotted as red circles (blue squares). Hollow symbols represent extracted barrier heights for switching distributions that do not agree well with the model.}
\end{figure}

\begin{figure*}
    \includegraphics[width=6.5 in,
    keepaspectratio=True]
    {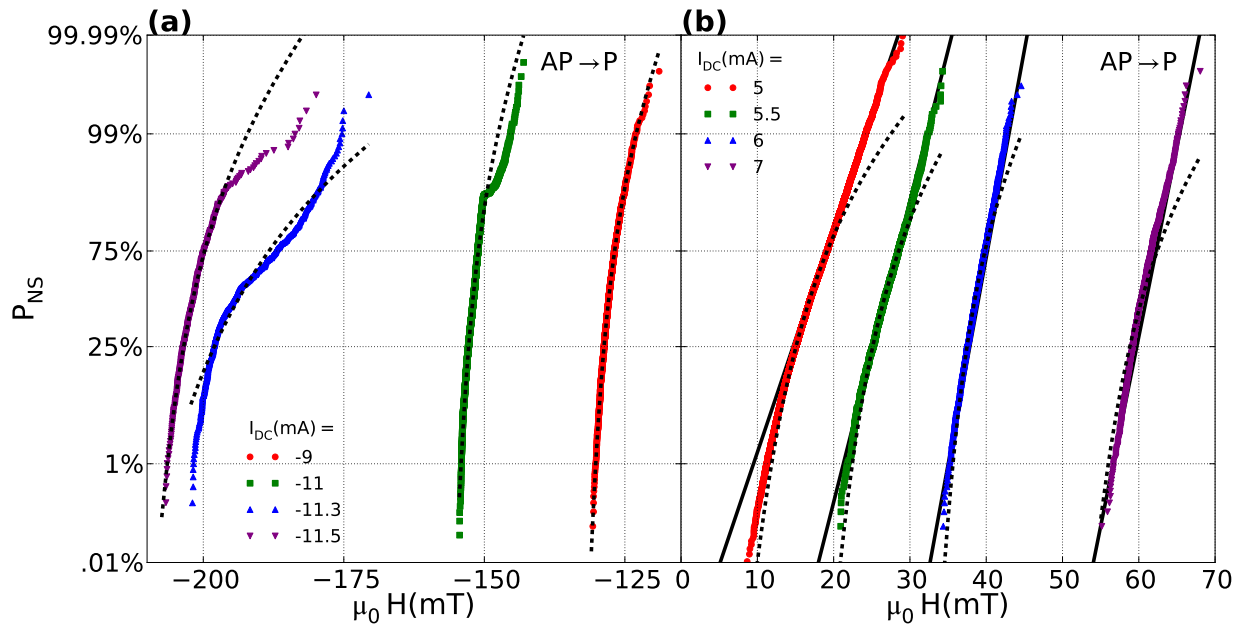}
  \caption{\label{fig:Deviations} Deviation of the switching field distributions from the predicted double-exponential shape at high currents. Switching distributions plotted on a Gauss quantile scale for (a) $\mathrm{I_{DC} = -9, -11, -11.3, -11.5 \, mA}$ reveal a kink for $\mathrm{I_{DC} \leq -11 \, mA}$ and (b)  $\mathrm{I_{DC} = 5, 5.5, 6, 7 \, mA}$ become apparently linear on a Gauss quantile scale. Broken lines in (a)\&(b) represent the best-fit thermal activation curve to the data. Solid lines in (b) represent the best-fit Normal distribution to the data.}
\end{figure*}

We extract the best-fit thermal stability factor for each switching field distribution, by fitting the thermal activation model to our data assuming a constant prefactor $\Gamma_0$ = 1~GHz and $v$ = 100 mT/s, the linear sweep rate of the applied external magnetic field. The best-fit parameter $\xi$ is shown as a function of applied current in Fig.~\ref{fig:Modified_Barrier}. The dissimilarity between $\xi$ for the AP$\rightarrow$P and P$\rightarrow$AP transitions at zero direct current is a common feature in perpendicularly magnetized SV devices and is related to an asymmetry caused by the polarizer dipole field.\cite{Gopman2012APL} We note that the trend for small currents is a monotonic decrease in $\xi$ for increasingly positive (negative) currents for the $\mathrm{AP \rightarrow P}$ ($\mathrm{P \rightarrow AP}$) transition, which is in good agreement with the effective barrier picture. On the other hand, as the current becomes increasingly negative (positive), the thermal stability for the $\mathrm{AP \rightarrow P}$ ($\mathrm{P \rightarrow AP}$) transition levels off. This is in contrast to the effective barrier model, which would predict a steadily increasing $\xi$ with current.\cite{Li2004} We denote these deviations from the model with hollow symbols in Fig.~\ref{fig:Modified_Barrier} to contrast with the switching field distributions that closely follow the monotonic trend line. We also apply hollow symbols for $\xi$ values extracted from switching field distributions that appreciatively deviate from the best model fit. As we will discuss below, this is also a consideration due to changes in the shape of the underlying switching field distribution for the higher current densities.

For sufficiently high current densities, the switching field distributions show clear deviations from the thermal distribution model, whose double-exponential distribution shape is evident in Fig.~\ref{fig:SFDs}. Figure~\ref{fig:Deviations}(a) illustrates the switching field distributions for large negative applied currents of -9,-11,-11.3, and -11.5~mA. At higher negative currents ($\mathrm{I_{DC}} \leq -11$~mA) the switching field distribution develops a kink (compared with $\mathrm{I_{DC}} = -9$~mA), which could indicate a cross-over between different competing reversal modes. These competing modes may involve excitation of the polarizer layer and are likely associated with the precessional modes typically seen at the edges of the bistable region of the state diagram, as in Fig.~\ref{fig:StateDiagram}(a). 

Figure~\ref{fig:Deviations}(b) illustrates the switching field distributions for large positive applied currents of +5,+5.5,+6, and +7~mA. At 5~mA, the shape of the switching field distribution first loses its curvature and the switching rate ($\mathrm{-dP_{NS}/dH}$) exceeds the thermal distribution model (dashed line) for fields below the median. Moreover, the switching rate for fields above the median is lower than predicted. This results in apparently linear distributions at 5.5, 6 and 7~mA when plotted on a Gauss quantile scale (compared to solid line), indicating that the switching rates are symmetric for a given deviation from the median switching field $\Delta = |H - H_0|$. 

In order to test whether any thermal process with multiple pathways can describe the data, we calculated switching field distributions assuming that multiple switching pathways are available for thermal activation. As it has been seen that spin-transfer torques can redistribute energy across fluctuation modes in a nanomagnet,\cite{Demidov2011, Slavin2009} competing fluctuation modes could be the origin of a distribution of switching pathways. We begin with a Gaussian distribution of switching rates $\Gamma_i$, each with their own energy barrier $\xi_i$. However, the energy barrier mainly determines the extent to which the $\mathrm{P_{NS}}$ curve bends above the knee, which makes it impossible for a distribution of these switching pathways to create a Gaussian switching field distribution. 

While the origin of the change in the shape of the switching field distributions is unclear, several factors may play a role at high currents. The nearly symmetric distributions in Fig.~\ref{fig:Deviations}(b) exhibit a lower switching rate for the low field, high-$\mathrm{P_{NS}}$ events than predicted by a thermal activation model, which may indicate that spin-transfer torques suppress the fluctuations that would result in thermal switching. Another possibility is that the large current densities exceeding $\mathrm{10^{11} A /m^2}$ may be driving the nanomagnet into an intermediate regime between thermally assisted reversal and deterministic (ballistic) switching in which neither a deterministic switching model nor a thermally-assisted switching model is valid.\cite{Bedau2010b} Switching field distributions at these higher currents may be reflecting this intermediate regime. 

In conclusion, we have tested the effective barrier model for spin-transfer assisted thermally activated reversal of spin-valve nanopillars with perpendicular magnetization. Although the effective temperature model catches the salient features of the average switching behavior seen in state diagram measurements, a closer investigation demonstrates a gradual deviation at the tails of a nanomagnet's switching distribution under increasing spin transfer torques. This shows that deep statistical measurements of the switching field combined with presentation on a Gauss quantile scale can reveal the onset and degree of deviation of the switching distributions from the thermal activation model. Clear deviations from the thermal switching model become apparent at currents exceeding the zero-field switching current. The deviations suggest a sudden change in the switching process and a breakdown in the validity of a model of thermally-induced switching. This could have significant impact on magnetic memory cells in which the bit write error rate in the tails may deviate significantly from the effective barrier model. Also, we demonstrate the saturating out of the barrier height for large current values. The origin of this as well as the effect of switching out of a dynamic state is not well understood. Nevertheless, our results demonstrate the need for additional investigations on the thermal stability of a nanomagnet under large spin transfer torques.

\section*{Acknowledgments}
This research was supported at NYU by NSF Grant Nos. DMR-1006575 and NSF-DMR-1309202, as well as the Partner University Fund (PUF) of the Embassy of France. Research at UL supported by ANR-10-BLANC-1005 ``Friends'', the European Project (OP2M FP7-IOF-2011-298060) and the Region Lorraine. Work at UCSD supported by NSF Grant No. DMR-1008654.

\bibliographystyle{apsrev}

\begin{thebibliography}{33}
\bibitem[{eta()}]{eta}
\bibinfo{note}{The arguments presented in this paper are valid for $1 \leq \eta
  \leq 2$, and only have a minor effect on the fitting parameters. $\eta= 1.5$
  is taken because this applies to cases in which a magnetic field is applied
  with a slight offset to the free layer easy axis.}
\end{thebibliography}

\end{document}